\documentclass[12pt,a4paper,twoside,english,british]{iopart}
\usepackage{mathptmx}
\usepackage[T1]{fontenc}
\usepackage[latin9]{inputenc}
\usepackage{amsbsy}
\usepackage{amstext}
\usepackage{amssymb}
\usepackage{graphicx}

\makeatletter

\pdfpageheight\paperheight
\pdfpagewidth\paperwidth

\usepackage{iopams}
\usepackage{setstack}

\@ifundefined{showcaptionsetup}{}{%
 \PassOptionsToPackage{caption=false}{subfig}}
\usepackage{subfig}
\makeatother

\usepackage{babel}
\begin{document}

\title{Demonstration of Free-space Reference Frame Independent Quantum Key
Distribution}

\author{J Wabnig}

\address{Nokia Research Center, Broers Building, 21 JJ Thomson Avenue, Cambridge
CB3 0FA, UK}

\author{D Bitauld}

\address{Nokia Research Center, Broers Building, 21 JJ Thomson Avenue, Cambridge
CB3 0FA, UK}

\author{H W Li}

\address{Nokia Research Center, Broers Building, 21 JJ Thomson Avenue, Cambridge
CB3 0FA, UK}

\author{A Laing }

\address{Centre for Quantum Photonics, H. H. Wills Physics Laboratory \& Department
of Electrical and Electronic Engineering, University of Bristol, BS8
1UB, UK}

\author{J L O\textquoteright{}Brien}

\address{Centre for Quantum Photonics, H. H. Wills Physics Laboratory \& Department
of Electrical and Electronic Engineering, University of Bristol, BS8
1UB, UK}

\author{A O Niskanen}

\address{Nokia Research Center, Broers Building, 21 JJ Thomson Avenue, Cambridge
CB3 0FA, UK}
\begin{abstract}
Quantum key distribution (QKD) is moving from research laboratories
towards applications. As computing becomes more mobile, cashless as
well as cardless payment solutions are introduced, and a need arises
for incorporating QKD in a mobile device. Handheld devices present
a particular challenge as the orientation and the phase of a qubit
will depend on device motion. This problem is addressed by the reference
frame independent (RFI) QKD scheme. The scheme tolerates an unknown
phase between logical states that varies slowly compared to the rate
of particle repetition. Here we experimentally demonstrate the feasibility
of RFI QKD over a free-space link in a prepare and measure scheme
using polarisation encoding. We extend the security analysis of the
RFI QKD scheme to be able to deal with uncalibrated devices and a
finite number of measurements. Together these advances are an important
step towards mass production of handheld QKD devices.
\end{abstract}

\submitto{\NJP}

\maketitle

\section{Introduction}

\begin{figure}[b]
\begin{centering}
\includegraphics[width=1\columnwidth]{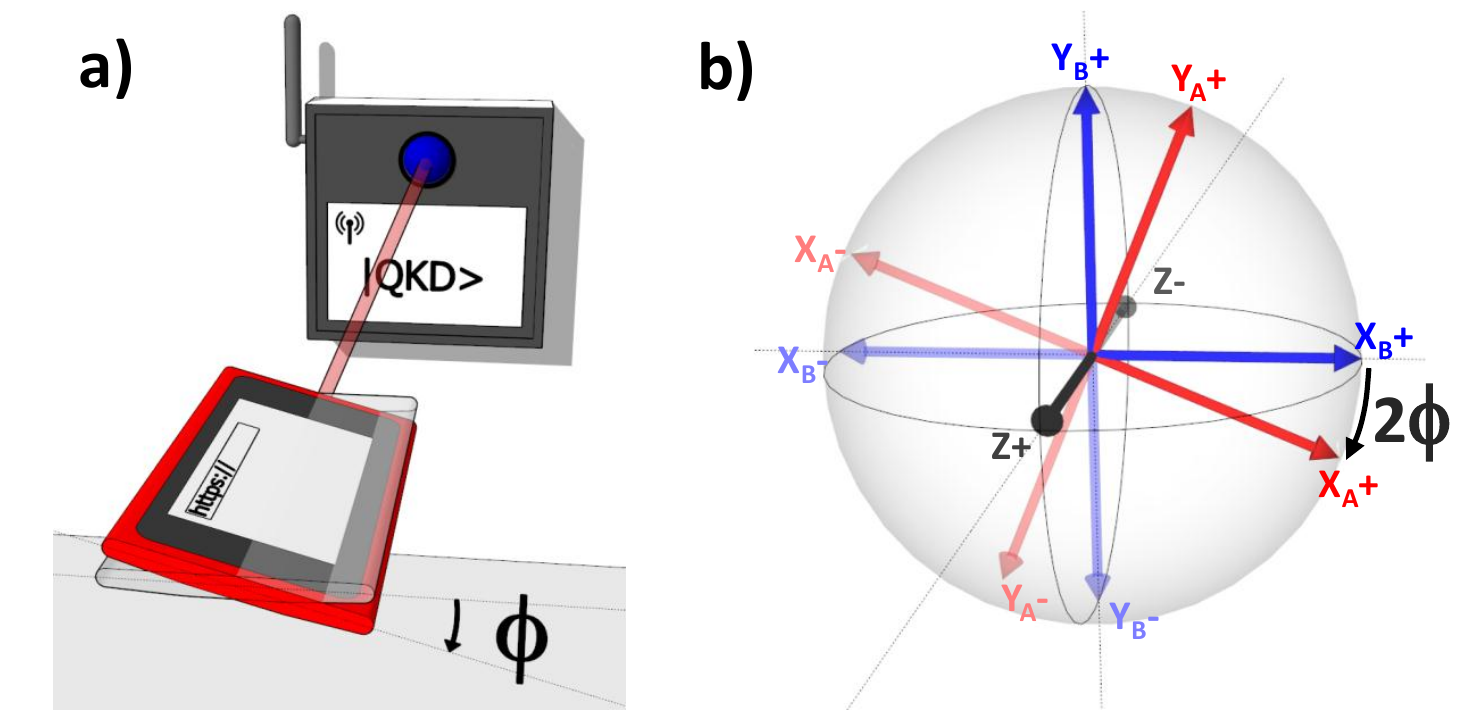}
\par\end{centering}

\caption{\label{Fig1}a) A mobile QKD terminal communicating with a stationary
QKD terminal. The mobile terminal is free to rotate. b) Poincaré sphere
representation of Alice and Bob\textquoteright{}s reference frames.
The Z basis (circular polarisation), represented by black vectors,
is shared by Alice and Bob and is used as the key basis. Alice\textquoteright{}s
X and Y bases (horizontal/vertical and diagonal/antidiagonal), represented
by red vectors, are aligned with the Poincaré sphere\textquoteright{}s
axes. Bob\textquoteright{}s X and Y bases, represented in blue, are
rotated with respect to Alice\textquoteright{}s. A rotation of the
terminal by the angle $\phi$ results in a rotation on the Poincaré
sphere by $2\phi$.}
\end{figure}

Quantum key distribution promises secure communications based not
on the hardness of a mathematical problem but on the laws of physics
\cite{2002RvMP...74..145G,Bennett1984,Ekert1991}. The main effort
in the development of QKD is directed towards long range communication,
mostly fibre-based \cite{2011NJPh...13l3001S,2011OExpr..1910387S,2009NJPh...11g5001P,2010ApPhL..96p1102D,2009OExpr..1711440D,2011AIPC.1363...35M}
as well as in free space \cite{2002NJPh....4...43H,2004NJPh....6...92A,2007NatPh...3..481U,2007PhRvL..98a0504S}.
A prospective new application of QKD is in securing short range line
of sight communications between a terminal and a handheld device (see
Fig~\ref{Fig1}a) \cite{2006SPIE.6399E..12G,2006NJPh....8..249D}.
Current mobile payment techniques, e.g. Near Field Communications,
have a range of security challenges including eavesdropping. Similar
considerations apply to securing Wi-Fi access points. We believe that
future handheld QKD systems can address these security challenges
and provide a high degree of wireless security.

One of the unique problems faced by handheld QKD is the fact that
the relative orientation of the emitter and the receiver is variable.
In previous works this problem has been addressed by using entanglement
\cite{PhysRevLett.93.220501} or by encoding information on angular
momenta \cite{2012NatCo...3E.961D}, which are invariant under rotation.
The Reference Frame Independent QKD scheme proposed in \cite{2010PhRvA..82a2304L}
does not require entanglement and allows the use of polarisation encoding
without the need for alignment of the qubit reference frames. In this
scheme the qubits are prepared and measured in three mutually unbiased
bases. Only one of those bases, on which the key is encrypted, needs
to be stable. The two other bases are allowed to drift slowly and
are used to estimate the security parameters of the quantum channels.
The requirement for one basis to be stable is met in most practical
implementations. In the case of free-space polarisation-based encoding,
the stable basis is the circular polarisation, which is rotation independent.
Fig~\ref{Fig1}b shows the bases used by the emitter (Alice) $X_{A}$,
$Y_{A}$, $Z_{A}$ and the receiver (Bob) $X_{B}$, $Y_{B}$, $Z_{B}$
for an undetermined relative orientation. The use of additional bases
compared to e.g. BB84 enables the reference frame independence of
the scheme.

In order to demonstrate the feasibility of such a protocol we implement
a prepare and measure scheme based on faint pulses and analyse the
security of the channel. In addition to the reference frame independence
implied by the protocol, we develop a theoretical analysis that allows
us to take into account deviations from the ideal perpendicular preparation
and measurement bases. This is similar in spirit to device independent
QKD (e.g. \cite{PhysRevLett.98.230501}). Although we cannot claim
to achieve full device independence, which generally requires entanglement,
we are able to deal with device imperfections and uncalibrated devices
within our selected model. In any QKD scheme the number of measurements
available for the parameter estimation step is finite. Parameters
obtained from the measurements are estimates with a non-zero variance,
which impacts the secure key fraction as discussed in \cite{2010NJPh...12l3019S,2006PhRvA..74d2340M,2008PhRvL.100t0501S}.
In our security analysis we take this into account to derive a secret
key fraction. In this article, section 2 describes the experimental
setup we used to implement a RFI protocol, while in section 3 we analyse
the security of the quantum channel including the influence of finite
size effects, imperfectly calibrated devices and mismatched detector
efficiencies. In section 4 we discuss our results.

\begin{figure*}[t]
\begin{centering}
\includegraphics[width=15cm]{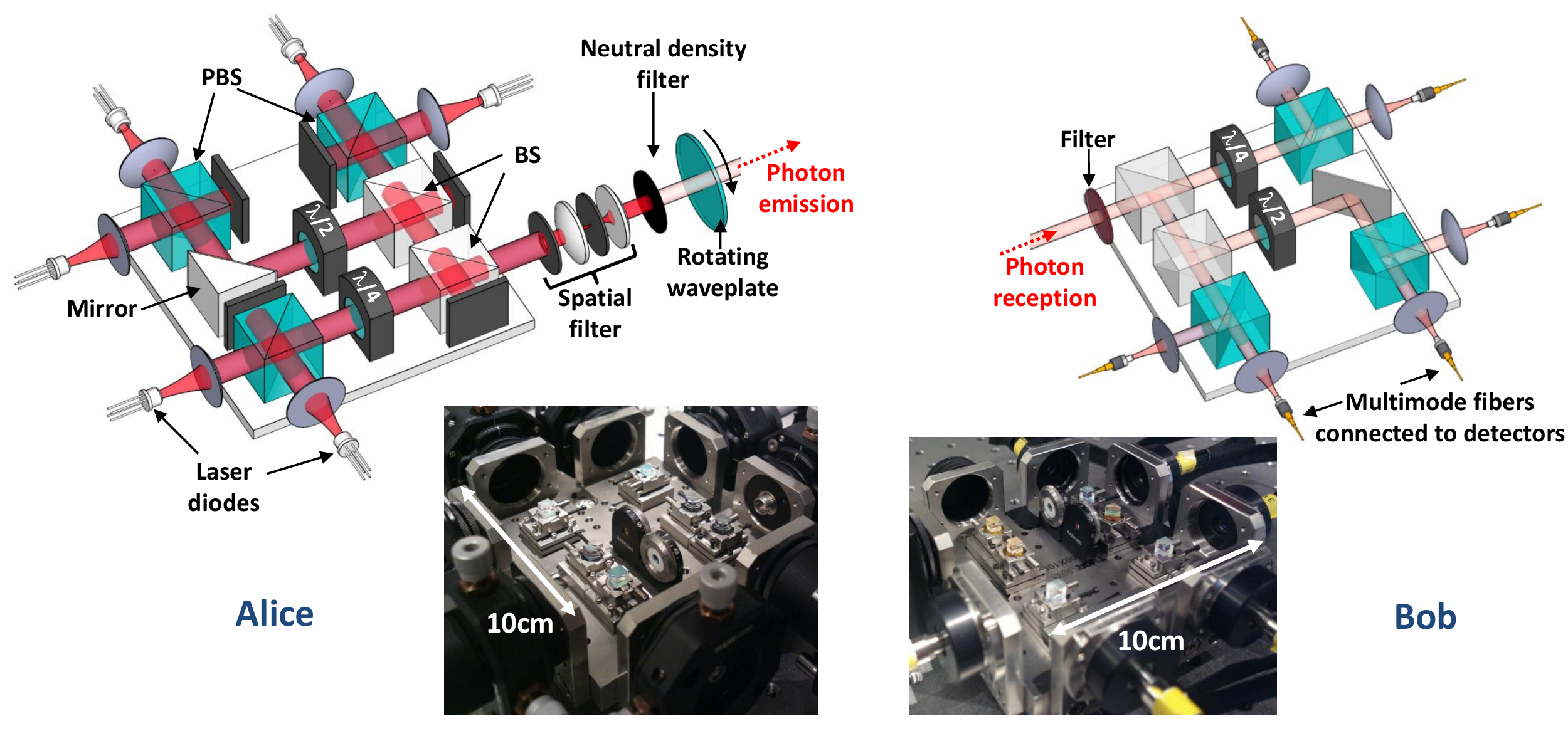}
\par\end{centering}

\caption{\label{Fig2}Experimental implementation of RFI QKD with the layout
of the emitter (Alice\textquoteright{}s device) and the receiver (Bob\textquoteright{}s
device). On Alice\textquoteright{}s side, unpolarised light is produced
by six laser diodes directed towards the two inputs of three polarising
beam splitters. The outputs of the polarising beam splitters are polarised
either horizontally or vertically depending on which laser is active.
One of the outputs passes through a $22.5\textdegree$ rotated $\lambda/2$
waveplate resulting in diagonal/antidiagonal polarisations. Another
PBS output is directed towards a $45\textdegree$ rotated $\lambda/4$
waveplate resulting in left/right circular polarisations. The last
polarising beam splitter output\textquoteright{}s polarisation is
left unchanged. The three resulting beams are combined with two non-polarising
beam splitters. The spatial profile of the combined beams is filtered
with a $1\,\textnormal{mm}$ pinhole, while the direction profiles
are filtered with a $5\,\mu\textnormal{m}$ pinhole between two lenses.
Finally, light passes through a spectral filter and a neutral density
filter reducing the intensity to single photon level. After the output
of the emitter, a rotating $\lambda/2$ waveplate is used to simulate
a rotation between the emitter and the receiver. The optical arrangement
of the receiver is similar to the emitter except that optical fibres
leading to the detectors are mounted instead of the lasers. Both Alice
and Bob\textquoteright{}s device are mounted on a $10\,\textnormal{cm}$
by $10\,\textnormal{cm}$ metal plate and the distance between the
two is slightly above $1\,\textnormal{m}$. The compactness of our
assembly can be seen in the insets.}
\end{figure*}

\section{Experimental setup}

In the prepare and measure RFI protocol, Alice needs to be able to
randomly pick a light polarisation state out of three different bases,
i.e. horizontal/vertical, antidiagonal/diagonal, left circular/right
circular, corresponding to X, Y and Z basis. This is implemented by
activating one of six 850~nm VCSEL laser diodes. Their respective
polarisations are separately engineered before their optical beams
are combined in a common mode. They are electrically driven by a pattern
generator with six individually controllable outputs. Each output
generates 0.5~ns pulses and their amplitudes are adjusted so that
the average photon number per pulse at Alice's output is the same
for all the lasers. Random bit sequences allowing only one of the
six lasers to be active every 4~ns (i.e., $250\times10^{6}$ pulses
per second) are programmed in the pattern generator's memory. Engineering
of the polarisations is performed by a set of polarising beam splitters
(PBS), and waveplates (see Fig~\ref{Fig2}). At first the polarisation
of the VCSELs is ill-defined but it is filtered by the polarisation
beam splitters with an extinction ratio of about 13~dB. The outputs
of the polarising beam splitters are polarised either horizontally
or vertically depending on which laser is active. One of the PBS output's
polarisation is left unchanged. The output of another PBS passes through
a $22.5\textdegree$ rotated $\lambda/2$ waveplate resulting in diagonal/antidiagonal
polarisations. The output of the last PBS is directed towards a $45\textdegree$
rotated $\lambda/4$ waveplate resulting in left/right circular polarisations.
The waveplate used in this experiment are achromatic but we estimate
their retardance at $850\,\textnormal{nm}$ to be 0.535 and 0.265
(+/-0.05) for the half and quarter waveplates respectively. Those
discrepancies together with the PBS extinction ratio are responsible
for the biasing of the bases that are discussed in the theoretical
section. After preparing the polarisations the three resulting beams
are combined with two non-polarising beam splitters. The spatial profile
of the combined beams is filtered with a $1\,\textnormal{mm}$ pinhole,
while the direction profiles are filtered with a $5\,\mu\textnormal{m}$
pinhole between two lenses. Finally, light passes through a neutral
density filter reducing the intensity down to below 0.05 photons per
pulse. After the output of the emitter, a rotating $\lambda/2$ waveplate
is used to simulate a rotation between the emitter and the receiver.
The photons then travel through free space for slightly more than
a meter to the measurement terminal. At the input of the measurement
device, a spectral filter with 10 nm passband and 0.7 maximum transmission
is used to reduce background noise. The optical arrangement of the
receiver is similar to the emitter except that it is used in reverse.
The common optical mode is divided in three beams by the non-polarising
beam splitters and waveplates are used before the PBSs to allow measurements
in the three bases. Thus, the measurement basis is passively selected
by the path of the photon, making it perfectly random. The six PBS
outputs are coupled with approximately 0.8 efficiency to multimode
optical fibres leading to single photon detectors. The photo detectors
are silicon avalanche photodiodes with 0.45 efficiency at 850~nm,
400 dark counts per second, a timing resolution of 600~ps and 50~ns
dead time. Their firing times as well as a clock pulse coming from
the pattern generator are continuously recorded by a counting card.
In post processing, we eliminate all the counts that happened when
another detector was still in its dead time, i.e. whenever two counts
happen within 60~ns. Bitrates could possibly be increased by using
detection events in other detectors during a given detector's deadtime,
but the implications on security are unexplored. The remaining number
of counts for a measurement duration of 1~s is approximately $2\times10^{6}$
out of the $250\times10^{6}$ weak pulses generated. The experiment
is performed in the dark and owing to spectral filtering background
noise is negligible compared to detector dark counts.

\begin{figure*}
\centering{}\subfloat[\label{Fig3a}]{\begin{centering}
\includegraphics[bb=1cm -2cm 627bp 474bp,width=4.5cm]{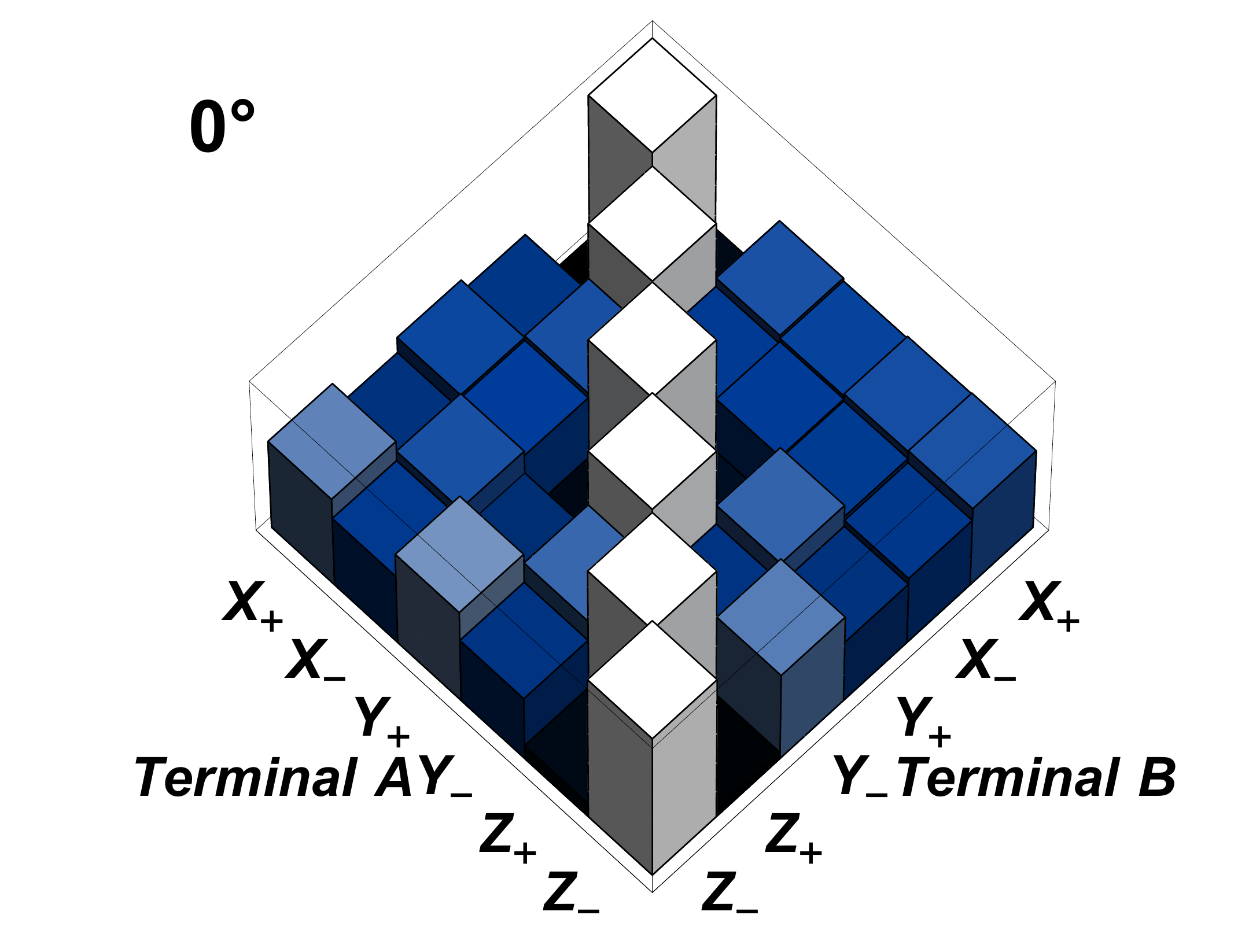}
\par\end{centering}

}\subfloat[\label{Fig3c}]{\centering{}\includegraphics[width=11cm]{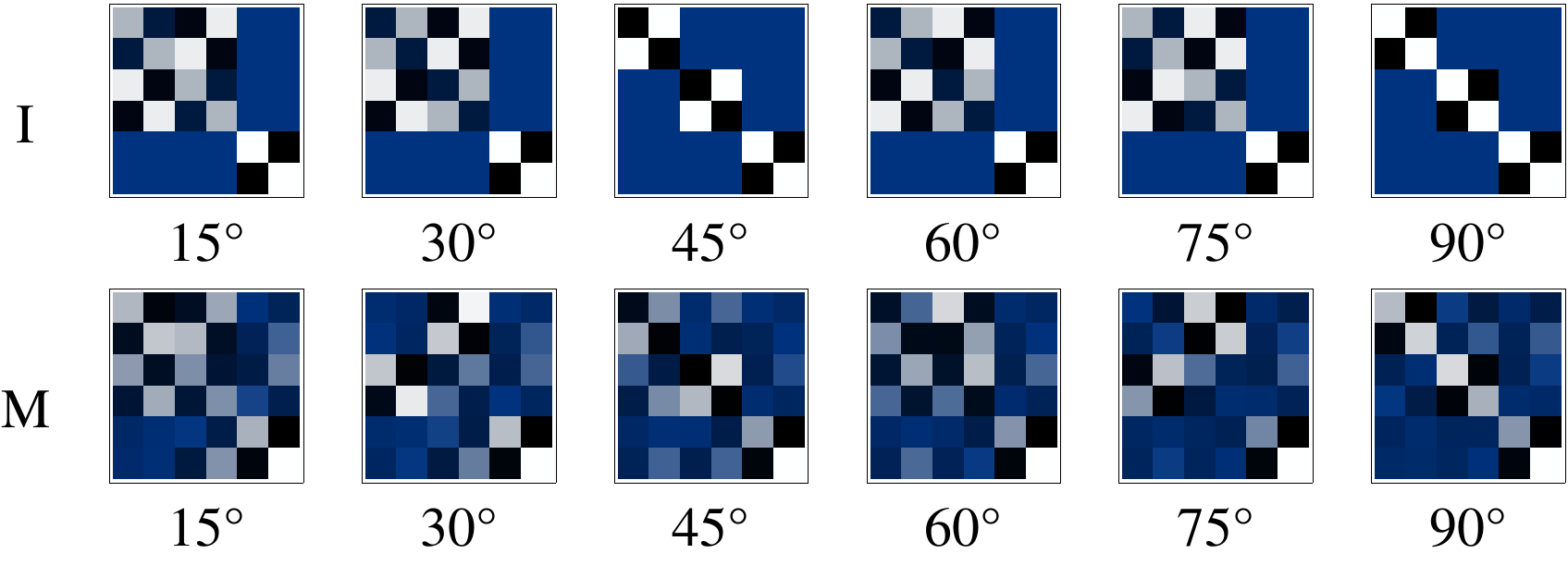}}\caption{\label{Fig3}Results of the normalised detector count matrix for the
implemented reference frame independent quantum key distribution protocol.
Each matrix square represents the number of counts on the specified
detector, when the specified polarisation was prepared. a) With the
two reference frames aligned, corresponding preparation and measurement
bases are almost completely correlated, with measurements between
mutually unbiased bases showing almost no correlation. This indicates
almost perfect transmission through the quantum channel. b) Comparison
of ideal (I) and measured data (M) for six further reference frames,
with angle of misalignment shown. While the counts contributing to
the correlator $C_{ZZ}$ remain largely independent of the rotation
angle, others contribute a periodic dependence to correlators $C_{XX}$,
$C_{XY}$, $C_{YX}$ and $C_{YY}$. Counts contributing to the non-zero
correlators $C_{XZ}$, $C_{YZ}$, $C_{ZX}$ and $C_{ZY}$ indicate
that the X, Y and Z directions are not perpendicular.}
 
\end{figure*}

Finally, by matching recorded detection times with the original random
patterns we can construct a 6 x 6 matrix (see Fig~\ref{Fig3a}) corresponding
to the number of times one of the six detectors fired when one of
the six lasers was active. The number of counts is maximal if the
emitting and receiving polarisations are the same, it is minimal if
they are opposite and it is approximately half of the maximum for
all the other polarisations. This matrix constitutes the raw data
of the security analysis. The counts in this matrix will be used for
two different purposes. Half of the counts when both the sending and
the measurement basis were Z are set aside as the raw key. The remaining
half with all other counts forms the basis of the parameter estimation.
The raw key length for each 1~s interval is approximately $2\times10^{5}$.
In order to demonstrate the robustness of our QKD scheme against reference
frame rotations, we insert a half wave plate on a rotating mount in
the free space optical path between the two terminals. Varying the
optic axis angle of the half wave plate simulates rotations of Alice's
device. For each angle, a measurement is taken similar to that presented
in Fig~\ref{Fig3a}. A comparison between measurement data and the
6 x 6 matrix predicted for ideal preparation and detection is shown
in Fig~\ref{Fig3c}.
 The matching detector count patterns in Fig~\ref{Fig3c} already
suggest that a quantum channel between the terminals is established,
but in order to claim the possibility of quantum key distribution
we need to analyse the results more rigorously.

\section{Secure Key Fraction}

The security of the RFI QKD protocol is discussed in \cite{2010PhRvA..82a2304L}:
Two parameters are estimated from the measurements: the quantum bit
error rate $Q$ and a rotation invariant quantity $C$ (see \ref{appB}).
While this analysis gives a good intuition why the scheme is reference
frame independent, it does not take into account factors that can
impact the key rate negatively in a real world setting, such as non-orthogonality
of the measurement directions or finite size effects. In this section
we derive the secure key fraction in the qubit subspace based on a
model of the device including imperfect calibration of preparation
and measurement bases, non uniform detector efficiencies and also
the effect of a finite number of measurements. In general the calculation
of the secret key fraction can be posed as an inference problem: given
the measurements and a device model what is the maximum amount of
information an eavesdropper can possess about the distributed key?
We first introduce a simplified expression for the secure key fraction,
then consider a model of our QKD system and minimise the secure key
fraction subject to constraints derived from the model and the measurements. 

The secure key fraction can be reduced from the ideal 100\% due to
leakage of information at two stages in the protocol \cite{2009RvMP...81.1301S}.
During the quantum stage information can leak to an eavesdropper,
while during error correction a certain minimum amount of information
needs to be exchanged over a classical channel. The two terms of the
following expression \cite{2005PhRvA..72a2332R} for the secret key
fraction reflect the impact of those two information leaks 
\begin{equation}
r=S(\chi_{A}|\rho_{E})-H(\chi_{A}|\chi_{B}),\label{KeyRate-1}
\end{equation}
where $S(\cdot|\cdot)$ denotes the conditional von Neumann entropy,
$H(\cdot|\cdot)$ the conditional Shannon entropy, $\chi_{A}$ are
the classical bit values at terminal $A$, $\chi_{B}$ are the classical
bit values at terminal $B$ and $\rho_{E}$ denotes the density matrix
of the eavesdropper. The first term describes the entropy of the key
bits $\chi_{A}$ given the eavesdropper's information, we will call
this ``usable entropy'', $S_{U}$. The second term is the minimum
necessary information to successfully perform error correction at
the Shannon limit. In any real world implementation this number has
to be multiplied by a factor larger than one to account for non-ideal
error correction schemes. The main objective of the parameter estimation
step is to find the minimum usable entropy given the constraints set
by the measurements. These constraints also take into account the
uncertainties associated with a finite number of performed measurements.
Additionally the parameter estimation step provides information about
the bit error rate, which is needed for error correction and thus
an estimate for the second term. In our analysis we assume that the
measurement results are produced by a two qubit density matrix $\rho_{AB}$
shared between parties $A$ and $B$. The usable entropy can be expressed
as (see \ref{appC})
\begin{equation}
S_{U}=S\left(\rho_{AB}||\mathcal{P}\rho_{AB}\right),
\end{equation}
where $S\left(A||B\right)$ denotes the relative entropy and the super-operator
$\mathcal{P}\rho=P_{0}^{A}\rho P_{0}^{A}+P_{1}^{A}\rho P_{1}^{A}$
with the projector on the bit values at terminal $A$ defined as $P_{0/1}^{A}=\frac{1}{2}\left(1\pm\sigma_{z}^{A}\right).$
To estimate the usable entropy we have to establish a model for our
quantum key distribution system. This consists of a model of the quantum
channel, embodied in the density matrix, along with a model of the
terminal devices. The secret key fraction can be obtained as the minimum
over all channel and device parameters that are consistent with the
performed measurements. For a finite number of measurements the constraints
will possess a statistical uncertainty that has to be taken into account
in the minimisation procedure. As we are looking for an upper bound
for the usable entropy it suffices to consider a simplified density
matrix (see \ref{appD}), guaranteed to give a lower usable entropy
than the full density matrix. The usable entropy for this density
matrix becomes
\begin{equation}
S_{U}(\lambda_{1},\lambda_{2})=1+\sum_{i}\eta_{i}\log_{2}\eta_{i}-h\left(2\eta_{1}\right),
\end{equation}
with the eigenvalues of the density matrix $\eta_{1}=\eta_{2}=1/4(1-\lambda_{1})$,
$\eta_{3}=1/4(1+\lambda_{1}-2\lambda_{2})$ and $\eta_{4}=1/4(1+\lambda_{1}+2\lambda_{2})$
and $h(x)=-x\log_{2}x-(1-x)\log_{2}(1-x)$. With perfectly calibrated
and aligned devices no additional model parameters have to be taken
into account. Imperfect measurement devices can lead to a large number
of additional parameters, such as non-orthogonalities in the preparation
and measurement bases and detector efficiencies, which can be collected
into the vector $\boldsymbol{\alpha}$ (for the details of the model
see \ref{appE}). The usable entropy is obtained as the minimum over
all parameters $\boldsymbol{\alpha}$, $\lambda_{1/2}$ 
\begin{equation}
S_{min}=\min_{\boldsymbol{\alpha},\lambda_{1},\lambda_{2}}S_{U}(\lambda_{1},\lambda_{2}).
\end{equation}
The parameters have to obey the constraints imposed by the observations
\begin{equation}
f_{i}(\boldsymbol{m})-\sigma\delta f_{i}(\boldsymbol{m})\le f_{i}\left[\boldsymbol{q}(\boldsymbol{\alpha},\lambda_{1},\lambda_{2})\right]\le f_{i}(\boldsymbol{m})+\sigma\delta f_{i}(\boldsymbol{m}),
\end{equation}
where $\boldsymbol{m}$ is a matrix containing all relevant detector
counts, $\boldsymbol{q}$ is the corresponding probability of observing
a detector count according to the device model (see \ref{appE}),
the $f_{i}$ are the functions defining the different constraints,
the $\delta f_{i}$ are their corresponding variances and $\sigma$
is chosen to give a certain probability that the estimated usable
entropy is too high. In our parameter estimation step we use a set
of 21 constraints (see \ref{appF}) consisting of 9 correlation functions
$C_{AB},\, A,B=X,Y,Z$, the six probabilities that a photon was prepared
in a certain polarisation direction $P_{A\pm},\, A=X,Y,Z$ and the
six probabilities to detect in a certain detector $D_{B\pm},\, B=X,Y,Z$.
For each function $f_{i}$ we can give the standard deviation $\delta f_{i}$
(see \ref{appF}). To obtain the secret key fraction we need to subtract
the observed relative entropy in the key basis from the usable entropy
\begin{equation}
r=S_{min}-h\left(\frac{1-C_{ZZ}+\sigma\delta C_{ZZ}}{2}\right).
\end{equation}
The full set of measurements enables us to calculate a reference frame
independent key rate. From the detector counts we can construct 9
correlation functions 
\begin{equation}
C_{AB}=\frac{m_{++}^{AB}+m_{--}^{AB}-m_{+-}^{AB}-m_{-+}^{AB}}{m_{++}^{AB}+m_{--}^{AB}+m_{+-}^{AB}+m_{-+}^{AB}},\, A,B=X,Y,Z,
\end{equation}
where the $m_{\pm\pm}^{AB}$ are the four different detector counts
given that the qubit was prepared in direction $A\pm$ and detected
in direction $B\pm$. In the case of orthonormal preparation and measurement
bases these can be directly identified with the qubit correlation
functions. A secret key fraction can be obtained as the minimum key
rate over all density matrices consistent with the observed correlators.
In the BB84 protocol \cite{1992PhRvL..68..557B,2005PhRvA..72a2332R}
only two correlators, e.g. $C_{XX}$ and $C_{ZZ}$, are used as constraints.
When the reference frames are rotated $C_{XX}$ will drop and so will
the secret key fraction. The RFI QKD scheme uses the four correlators
$C_{XX}$, $C_{XY}$, $C_{YX}$, $C_{YY}$ as well as $C_{ZZ}$. While
each individual correlator may decrease under reference frame rotations,
the combination $C=C_{XX}^{2}+C_{XY}^{2}+C_{YX}^{2}+C_{YY}^{2}$ ideally
stays constant, enabling reference frame independent secret key fractions.
In order to treat imperfect preparation and measurement we have to
depart from the assumption of orthonormal bases and include a more
detailed detector model. Each preparation and each detector are associated
with a direction on the Poincaré sphere, given by a unit vector. Since
we aim to use the z-basis for the key bits we can identify two preparation
directions with the $\pm z$ direction on the Poincaré sphere, without
overestimating the secret key fraction. Each direction is parametrised
by two variables (e.g. azimuth and polar angle), resulting in a total
of 20 free parameters (see \ref{appE}). Different absorption may
occur in the preparation channels, and similarly detectors may have
different efficiencies. With six possible preparation directions and
six possible detection direction this adds another set of 12 parameters.
The quantum channel can be represented by a two parameter two qubit
density matrix in a simplification over the more commonly employed
Bell diagonal density matrix (see \ref{appD}). This results in a
model with 34 free parameters. The secret key fraction is then obtained
as the minimum over all parameters that fulfil the constraints imposed
by the measurements, e.g. from the correlators $C_{AB}$; in our case
a set of 21 constraints. For the number of detector counts approaching
infinity the constraints are equalities, but for a finite number of
observations the value can lie within an interval determined by the
number of counts and the desired uncertainty. A small number of counts
will lead to larger uncertainty in the correlation function and therefore
to lower results for the secret key fraction.

\section{Results and Discussion}

\begin{figure}
\begin{centering}
\includegraphics[width=0.8\columnwidth]{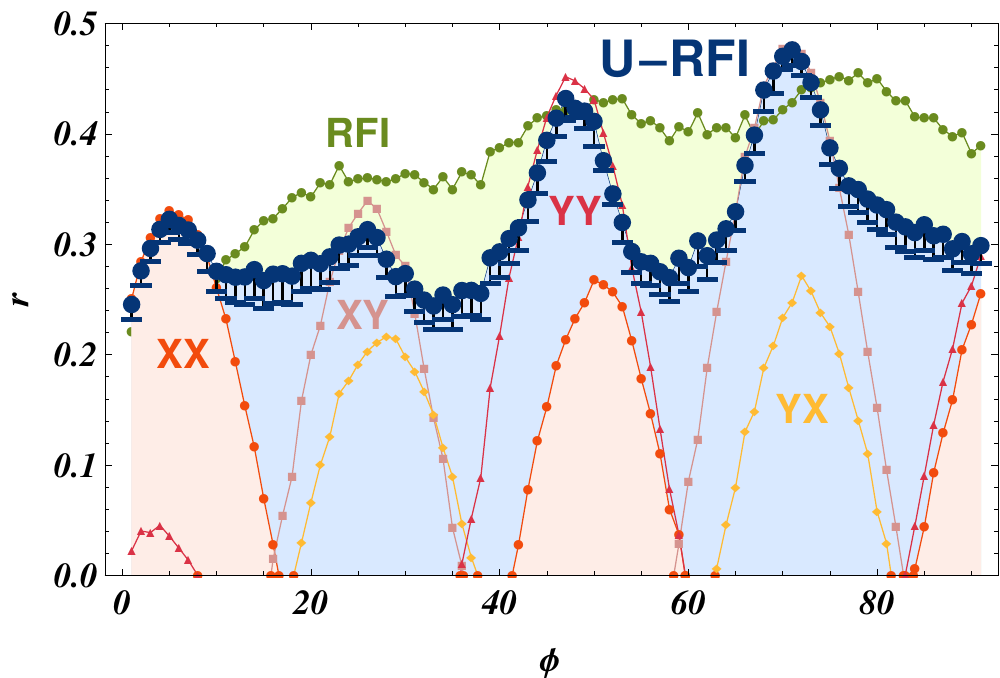}
\par\end{centering}

\caption{\label{Fig4}Keyrates as a function of the physical waveplate rotation
angle for two reference frame independent and 4 different BB84 protocols.
U-RFI labels the reference frame independent keyrate assuming uncalibrated
devices. It is obtained from numerical minimisation over the full
set of model parameters subject to constraints imposed by the measurements.
The blue dots correspond to the minimised key rate neglecting the
variance coming from the finite number of measurements. The blue bars
indicate the result obtained by allowing for an uncertainty in the
estimate of the constraints of three times the standard deviation.
The remaining curves are obtained from the same data but with a simplified
model, assuming orthogonality within bases and total perpendicularity
between bases. RFI labels the keyrate for the reference frame independent
protocol as outlined in \cite{2010PhRvA..82a2304L}. XX, XY, YX and
YY label the keyrates for the BB84 protocol using the respective correlator
pairs $C_{ZZ}$ and $C_{XX}$, $C_{ZZ}$ and $C_{XY}$, and so on. }
\end{figure}
Our main result is that the secret key fraction remains above $0.2$
for all rotation angles even in the presence of device imperfections,
see U-RFI in Fig~\ref{Fig4}. To better understand the calculated
keyrate we make a comparison with established protocols: the BB84
protocol and the RFI QKD protocol as proposed by Laing, Scarani, Rarity
and O'Brien \cite{2010PhRvA..82a2304L}. For the sake of comparison
we assume that in these protocols both preparation bases and measurement
bases are perfect, i.e. the measured correlation functions are indeed
the qubit correlation functions of the underlying density matrix.
A summary of the calculation of the BB84 and RFI QKD key rates can
be found in \ref{appA} and \ref{appB}. We clearly see in Fig~\ref{Fig4}
that while the BB84 key rate drops to zero for certain angles the
RFI QKD key rate remains non-zero for all rotation angles. The plot
also shows that for certain angles the RFI secret key fraction estimate
is too optimistic compared to the U-RFI rates. This happens due to
non-orthogonal measurement directions, which are not taken into account
in the RFI QKD security analysis. The distinct peaks in the secret
key fraction coincide with the peaks visible in the secret key fraction
calculated for the BB84 protocols, as outlined above. This can be
understood in the following way: The secret key fraction we calculate
depends on the correlators $C_{AB},\, A,B=X,Y,Z$. The correlator
$C_{ZZ}$ is linked to the qubit error rate. The correlators $C_{XX}$,
$C_{XY}$, $C_{YX}$, $C_{YY}$ are related to monitoring the eavesdropping.
If the absolute value of one of these correlators is large then eavesdropping
in the key basis is small, even if other correlators are relatively
small. This also makes the secret key fraction larger then the RFI
keyrate for some angles. In the region between the peaks no individual
correlator is large and only their combination gives a positive secret
key fraction. In the presence of alignment errors some of the contributing
correlators may be small and as a result the secret key fraction drops. 

During each transmission of one second we set aside approximately
$2\times10^{5}$ raw key bits. After obtaining the correct secret
key fraction from the parameter estimation step we can perform the
privacy amplification step by applying the appropriate 2-universal
hash function (e.g. a Toeplitz matrix \cite{674930}) to the raw key
and obtain a secure key. With a secret key fraction of around $0.25$
we arrive at an approximate key rate of $5\times10^{4}\,\textnormal{s}^{-1}$.
In the current implementation of our scheme potential security loop
holes exist due to side channels \cite{2009NJPh...11f5001N}, the
most obvious one coming from the use of six different lasers in the
preparation stage. While this loophole can be plugged by using a single
laser and a 1 x 6 optical switch further analysis of the impact of
side channels on the key rate is required (e.g. the effect of photon
number statistics \cite{Lo2005} or detector characteristics \cite{2011PhRvL.107k0501J}
on our scheme). 

Nevertheless we are able to give an estimate of the impact of photon
number statistics on the key rate in our scheme. Without employing
decoy states we have to assume that in pulses which contain more than
one photon and which are contributing to the raw key all information
is lost to the eavesdropper. Our photon source generates weak coherent
pulses with a rate of $250\,\textnormal{MHz}$ and an average photon
number of 0.05 per pulse. In order to assess the impact of a hypothetical
photon number splitting attack we have to characterise the quantum
channel. An important quantity is the \emph{accessible loss} \cite{2002NJPh....4...44L}
in the quantum channel, i.e. the part of the loss that is, in principle,
under the control of the eavesdropper. Due to the short range free
space transmission no absorption occurs between Alice and Bob. A small
amount of absorption (coupling efficiency$\approx0.8$) can be attributed
to the mode mismatch between sender and receiver and could be manipulated
by an eavesdropper. All other losses occur in the Bob device and are
\emph{inaccessible losses}. Starting from a $250\,\textnormal{MHz}$
pulse rate and 0.05 photons per pulse we arrive at a raw single photon
rate of $12.5\,\textnormal{MHz}$. The recorded photon count rate
is $2\,\textnormal{MHz}$, giving a total absorption of $\mbox{\ensuremath{\eta}=0.16}$.
The accessible part of the absorption is $\eta_{A}=0.8$, while the
inaccessible part, containing absorption in filters, losses in optical
components and finite detector efficiency contributes $\eta_{I}=0.2$,
so that $\eta=\eta_{A}\eta_{I}$. The rate of two photon pulses according
to Poisson statistics is $0.3\,\textnormal{MHz}$. The remaining single
photons after a photon number splitting attack will still experience
the inaccessible losses, so that the total number of recorded detector
clicks associated with two photon pulses is $60\,\textnormal{kHz}$.
Of these photons only a fraction contributes to the raw key. For our
setup that fraction is $0.1$. For a key distribution lasting one
second we estimate that an eavesdropper can at most have information
about $6000$ key bits out of $2\times10^{5}$ key bits, or the secure
key fraction has to be reduced by approximately 0.03, due to the probabilistic
nature of the photon source, only slightly modifying our secret key
fraction.

In summary, we have shown the feasibility of reference frame independent
key distribution using off the shelf bulk optical components in a
compact assembly. Our setup uses passive components and can readily
be transferred to a miniaturised version using integrated optics on
a chip \cite{2008Sci...320..646P,2009NaPho...3..687O}. Our analysis
employed in the parameter estimation step obviates the need for precise
alignment of the preparation and measurement qubit bases. The achievable
key rates are sufficient to distribute hundreds of 256 bit keys within
1 second. In order to realise a handheld QKD device additional functionality,
like steering of the emitted photons towards a receiver needs to be
implemented. Possible solutions include movable mirrors, movable lenses
or phased arrays. Overall, the work presented here paves the way for
mass-produced handheld QKD devices.

\appendix

\section{Keyrate for BB84\label{appA}}

For the calculation of the keyrate for the BB84 protocol we follow
\cite{2005PhRvA..72a2332R}, with the difference that we use two parameters
for the quantum channel, the correlation functions $C_{XX}$ and $C_{ZZ}$
. The secure key fraction for the BB84 protocol is given by
\begin{equation}
r=\min_{\eta_{i}}\left(1+\sum_{i=1}^{4}\eta_{i}\log_{2}\eta_{i}\right)
\end{equation}
under the constraints 
\begin{equation}
C_{XX}=1-2\eta_{2}-2\eta_{4}
\end{equation}

\begin{equation}
C_{ZZ}=1-2\eta_{3}-2\eta_{4}
\end{equation}
\begin{equation}
\sum_{i=1}^{4}\eta_{i}=1.
\end{equation}
This can be solved analytically to give
\begin{equation}
r=1+z_{+}x_{+}\log_{2}\left(z_{+}x_{+}\right)+z_{+}x_{-}\log_{2}\left(z_{+}x_{-}\right)+z_{-}x_{+}\log_{2}\left(z_{-}x_{+}\right)+z_{-}x_{-}\log_{2}\left(z_{-}x_{-}\right)
\end{equation}
with
\begin{equation}
x_{\pm}=\frac{1\pm C_{XX}}{2},\quad z_{\pm}=\frac{1\pm C_{ZZ}}{2}
\end{equation}
Alternatively other correlators than $C_{XX}$ can be used to obtain
a keyrate. In the case of misaligned reference frames the correlators
$C_{XY}$ or $C_{YX}$ may give a more favourable keyrate.

\section{RFI QKD\label{appB}}

Reference frame independent quantum key distribution was introduced
in \cite{2010PhRvA..82a2304L}. The scheme solves the problem of aligning
reference frames between two partners in a quantum key distribution
protocol if one stable direction exists (e.g. in polarisation encoding
physical rotation does not influence the circular polarisation direction).
The stable direction is used for encoding the qubits, while perpendicular
directions are used for parameter estimation. It is sufficient to
consider two quantities constructed from the qubit correlation functions:
the bit error rate
\begin{equation}
Q=\frac{1-C_{ZZ}}{2}
\end{equation}
and the quantity
\begin{equation}
C=C_{XX}^{2}+C_{XY}^{2}+C_{YX}^{2}+C_{YY}^{2}.
\end{equation}
with the qubit correlation functions given by
\begin{equation}
C_{AB}=\textnormal{Tr}\left(\sigma_{A}\sigma_{B}\rho\right),
\end{equation}
where $\rho$ is the two qubit density matrix. Under rotations around
the z axis the Pauli matrices transform $\sigma_{x}\rightarrow\sigma_{x}\cos\alpha+\sigma_{y}\sin\alpha$
and $\sigma_{y}\rightarrow\sigma_{y}\cos\alpha-\sigma_{x}\sin\alpha$,
with the rotation angle given by $\alpha$. One can see that the quantity
$C$ stays invariant under these kind of rotations. The secret key
fraction can be shown to be a function of $C$ and $Q$. For $Q\lesssim0.159$
the secret key fraction becomes
\begin{equation}
r=1-h(Q)-(1-Q)h\left(\frac{1+u}{2}\right)-Qh\left(\frac{1+v}{2}\right)
\end{equation}
with $h(x)=-\log_{2}x-(1-x)\log_{2}(1-x)$ the Shannon entropy function
and $u=\min(\sqrt{C/2}/(1-Q),1)$ and $v=\sqrt{C/2-(1-Q)^{2}u^{2}}/Q$.

\section{\label{appC}Conditional Entropy decreases under Projective Measurement}

A quantum communication channel between parties Alice ($A$) and Bob
($B$) can be described by the density matrix $\rho_{AB}$. In general
this density matrix will not be pure due to noise, errors and the
action of an eavesdropper. In order to prove security of the communication
channel we have to assume all deviations from the ideal case are due
to an eavesdropper. The combined state of $\rho_{AB}$ and an eavesdropper
can then be expressed as the pure state
\begin{equation}
|\psi>_{ABE}=\sum_{i=1}^{4}\sqrt{\lambda_{i}}|\Phi_{i}>_{AB}|\nu_{i}>_{E},
\end{equation}
where the $\lambda_{i}$ are the eigenvalues of $\rho_{AB}$, $|\Phi_{i}>_{AB}$
are the corresponding eigenfunctions and $|\nu_{i}>$ are an orthogonal
basis for the state of the eavesdropper. Note that the dimension of
the Hilbert space of the eavesdropper equals the dimensions of $\rho_{AB}$.
The density matrices in the subspaces can then be obtained as
\begin{equation}
\rho_{AB}=\textnormal{Tr}_{E}\left(|\psi>_{ABE}<\psi|_{ABE}\right)
\end{equation}
and 
\begin{equation}
\rho_{E}=\textnormal{Tr}_{AB}\left(|\psi>_{ABE}<\psi|_{ABE}\right).
\end{equation}
The reduced density matrix of the eavesdropper equals the reduced
density matrix of Alice and Bob up to a unitary transformation
\begin{equation}
\rho_{E}=U\rho_{AB}U^{\dagger}.
\end{equation}
Let us consider the entropy of the key bits $\chi_{A}$ given that
the eavesdropper knows the state of the system $\chi_{E}$, $S(\chi_{A}|\chi_{E})$.
We can rewrite the conditional entropy as
\begin{equation}
S(\chi_{A}|\rho_{E})=S(\chi_{A})+S(\chi_{E}|\chi_{A})-S(\rho_{E}).
\end{equation}
With $X$ being the classical value of Alice's qubit we can write
\begin{equation}
S(\rho_{E}|\chi_{A})=\sum_{x=0,1}p_{x}S(E_{x})
\end{equation}
If Alice prepares key bit 0 with probability $p_{0}$ and key bit
one with probability $p_{1}$ we can write the corresponding density
matrices $E_{0}$, $E_{1}$ with the help of the projection operators
\begin{equation}
P_{x}=|x>_{A}<x|_{A},\quad x=0,1
\end{equation}
as
\begin{equation}
E_{x}=\textnormal{Tr}_{AB}\left(P_{x}|\psi>_{ABE}<\psi|_{ABE}\right)
\end{equation}
or 
\begin{equation}
E_{x}=\frac{1}{p_{x}}\sqrt{\rho_{AB}}P_{x}\sqrt{\rho_{AB}}.
\end{equation}
Since unitary transformations do not change the entropy we can replace
$\rho_{AB}$ with $\rho_{E}$ in the calculation of entropies 
\begin{equation}
S\left(\frac{1}{p_{x}}\sqrt{\rho_{AB}}P_{x}\sqrt{\rho_{AB}}\right)=S\left(\frac{1}{p_{x}}\sqrt{\rho_{E}}P_{x}\sqrt{\rho_{E}}\right).
\end{equation}
We can use 
\begin{equation}
\textnormal{Tr}\left[\sqrt{\rho}P\sqrt{\rho}F(\sqrt{\rho}P\sqrt{\rho})\right]=\textnormal{Tr}\left[P\rho PF(P\rho P)\right],
\end{equation}
where $F$ is any function and we used 
\begin{equation}
F(\sqrt{\rho}P\sqrt{\rho})=\sum_{n}c_{n}(\sqrt{\rho}P\sqrt{\rho})=\sum_{n}\sqrt{\rho}P\left(P\rho P\right)^{n-1}P\sqrt{\rho}.
\end{equation}
We can therefore write
\begin{equation}
S(\rho_{E}|\chi_{A})=\sum_{x=0,1}p_{x}S(\frac{1}{p_{x}}P_{x}\rho_{E}P_{x}).
\end{equation}
Since the $P_{x}$ project onto orthogonal subspaces we can use (see
Nielsen and Chuang, p.518 \cite{citeulike:541803}) to write 
\begin{equation}
S(\rho_{E}|\chi_{A})=S\left(\sum_{x=0,1}P_{x}\rho_{E}P_{x}\right)-S(\chi_{A}).
\end{equation}
The equality holds if the projectors $P_{x}$ project onto orthogonal
subspaces and acts as an upper bound on the usable entropy otherwise.
The conditional entropy therefore becomes
\begin{equation}
S(\chi_{A}|\rho_{E})=S\left(\sum_{x=0,1}P_{x}\rho_{E}P_{x}\right)-S(\chi_{E}).
\end{equation}
Introducing the projective measurement
\begin{equation}
\mathcal{P}\rho=\sum_{x}P_{x}\rho P_{x}
\end{equation}
we can write 
\begin{equation}
S(\chi_{A}|\rho_{E})=S\left(\mathcal{P}\rho_{E}\right)-S(\rho_{E}).
\end{equation}
Since $\mathcal{P}$ is a projective measurement this simplifies further
to the relative entropy

\begin{equation}
S(\chi_{A}|\rho_{E})=S\left(\rho_{E}||\mathcal{P}\rho_{E}\right)=S\left(\rho_{AB}||\mathcal{P}\rho_{AB}\right).
\end{equation}
We now want to investigate how a further projective measurement influences
the relative entropy. Introducing 
\begin{equation}
\mathcal{L}\rho=\sum_{i}L_{i}\rho L_{i},
\end{equation}
with the projection operators $L_{i}$. We consider the relative entropy
before and after the projective measurement $\mathcal{L}$
\begin{equation}
S\left(\rho_{AB}||\mathcal{P}\rho_{AB}\right)-S\left(\mathcal{L}\rho_{AB}||\mathcal{P}\mathcal{L}\rho_{AB}\right).
\end{equation}
For commuting operations 
\begin{equation}
\mathcal{PL}=\mathcal{LP}
\end{equation}
and using that the relative entropy of two density matrices decreases
or stays the same under completely positive trace preserving maps
\cite{1975CMaPh..40..147L} we obtain
\begin{equation}
S\left(\rho_{AB}||\mathcal{P}\rho_{AB}\right)-S\left(\mathcal{L}\rho_{AB}||\mathcal{L}\mathcal{P}\rho_{AB}\right)\geq0
\end{equation}
if the operations $\mathcal{L}$ and $\mathcal{P}$ commute.

\section{\label{appD}Simplified density matrices for parameter estimation}

We can use the inequality above to simplify the model density matrix
for parameter estimation. Starting from the full two qubit density
matrix with 15 free parameters we can apply a series of projective
measurements,
\begin{equation}
\mathcal{L}=\mathcal{L}_{2}\mathcal{L}_{1}
\end{equation}
to arrive at a simpler density matrix with fewer parameters that is
guaranteed to give a smaller value for the usable entropy. The individual
projective measurement super-operators are 
\begin{equation}
\mathcal{L}_{i}=\lim_{t\rightarrow\infty}e^{-t\mathcal{K}_{i}}
\end{equation}
and the corresponding Liouvillian 
\begin{equation}
\mathcal{K}_{i}\rho=\left\{ \Sigma_{i}\Sigma_{i},\rho\right\} -2\Sigma_{i}\rho\Sigma_{i},\quad\Sigma_{1}=\sigma_{z}^{A}-\sigma_{z}^{B},\Sigma_{2}=\sigma_{x}^{A}\sigma_{x}^{B}.
\end{equation}
If we use the z basis for the key bits then the measurement operator
becomes 
\begin{equation}
\mathcal{P}\rho=P_{0}^{A}\rho P_{0}^{A}+P_{1}^{A}\rho P_{1}^{A}
\end{equation}
 with the projector on the bit values at terminal $A$ defined as
\begin{equation}
P_{0/1}^{A}=\frac{1}{2}\left(1\pm\sigma_{z}^{A}\right)
\end{equation}
It can be shown that
\begin{equation}
\mathcal{P}\mathcal{L}_{i}\rho-\mathcal{L}_{i}\mathcal{P}\rho=0
\end{equation}
for all density matrices $\rho$. After applying all the projective
measurements to a general density matrix the simplified density matrix
takes the form 
\begin{equation}
\rho=\frac{1}{4}\left(\begin{array}{cccc}
1+\lambda_{1} & 0 & 0 & 2\lambda_{2}\\
0 & 1-\lambda_{1} & 0 & 0\\
0 & 0 & 1-\lambda_{1} & 0\\
2\lambda_{2} & 0 & 0 & 1+\lambda_{1}
\end{array}\right)\label{simplerho}
\end{equation}
This density matrix can now be used in models for parameter estimation.

\section{\label{appE}Device model }

We want to construct a model for the probability of registering a
detector count in the detector measuring in basis $B$, direction
$V$, provided the bit was prepared in basis $A$, direction $U$,
$p_{UV}^{AB}$. The probability of observing a count in a detector
given a certain preparation is proportional to 
\begin{equation}
q_{UV}^{AB}=t_{AU}^{1}t_{BV}^{2}\textnormal{Tr}\left(\hat{P}_{AU}\hat{P}_{BV}\rho\right),
\end{equation}
where $t^{1}$ and $t^{2}$ account for the preparation and detection
efficiencies. The projectors $\hat{P}_{AU},\,\hat{P}_{BV}$ are along
a certain preparation or measurement direction, given by a unit vector
$\boldsymbol{n}_{AX}$ or $\boldsymbol{r}_{BY}$
\begin{equation}
\hat{P}_{AU}=\frac{1}{2}\left(1+\boldsymbol{n}_{AU}\cdot\boldsymbol{\sigma}_{A}\right),\quad\hat{P}_{BV}=\frac{1}{2}\left(1+\boldsymbol{r}_{BV}\cdot\boldsymbol{\sigma}_{B}\right),
\end{equation}
where $\boldsymbol{\sigma}_{A/B}$ is a vector of Pauli matrices in
the respective local basis. Using the simplified density matrix Eq.~(\ref{simplerho})
we obtain
\begin{equation}
q_{UV}^{AB}=\frac{t_{AU}^{1}t_{BV}^{2}}{4}\left[1+\sum_{ij}n_{i}^{AU}r_{j}^{BV}\textnormal{Tr}\left(\hat{\sigma}_{i}^{A}\hat{\sigma}_{j}^{B}\rho\right)\right]
\end{equation}
with
\begin{equation}
\textnormal{Tr}\left(\hat{\sigma}_{i}^{A}\hat{\sigma}_{j}^{B}\rho\right)=\delta_{ij}l_{i}=\Lambda_{ij},\quad\boldsymbol{l}=(\lambda_{2},\lambda_{2},\lambda_{1})
\end{equation}
and therefore
\begin{equation}
q_{UV}^{AB}=\frac{t_{AU}^{1}t_{BV}^{2}}{4}\left(1+\boldsymbol{n}_{AU}\cdot\Lambda\cdot\boldsymbol{r}_{BV}\right).
\end{equation}
so that the probabilities become

\begin{equation}
p_{UV}^{AB}=\frac{q_{UV}^{AB}}{\sum_{ABUV}q_{UV}^{AB}}.
\end{equation}
For our measurement setup there are six distinct preparation directions
$\boldsymbol{n}_{AU},\quad A=X,\, Y,\, Z,\quad U=\pm$, with two free
parameters, the azimuthal and polar angle, specifying each direction.
The z-direction is identified with the qubit directions at terminal
A and therefore we fix $\boldsymbol{n}_{Z\pm}=(0,0,\pm1)$. Similarly
there are six measurement directions $\boldsymbol{r}_{BV},\quad B=X,\, Y,\, Z,\quad V=\pm$.

\section{\label{appF}Detector counts and correlation functions}

The calculation of the secret key fraction is based on a minimisation
subject to a number of constraints. In our implementation we use a
total of 21 constraints. Each constraint and its corresponding standard
deviation can be calculated from the raw detector counts obtained
in the parameter estimation process. The first set of constraints
is given by the 9 correlation functions
\begin{equation}
C_{AB}(\boldsymbol{m})=\frac{m_{++}^{AB}+m_{--}^{AB}-m_{+-}^{AB}-m_{-+}^{AB}}{m_{++}^{AB}+m_{--}^{AB}+m_{+-}^{AB}+m_{-+}^{AB}},\, A,B=X,Y,Z
\end{equation}
Additional 6 constraints are the probabilities for preparing a certain
direction
\begin{equation}
P_{AU}(\boldsymbol{m})=\frac{\sum_{BV}m_{UV}^{AB}}{N_{0}},\quad N_{0}=\sum_{ABUV}m_{UV}^{AB},
\end{equation}
where $N_{0}$ is the total number of detector counts. Further 6 constraints
are the probabilities for detecting a certain direction

\begin{equation}
D_{BV}(\boldsymbol{m})=\frac{\sum_{AU}m_{UV}^{AB}}{N_{0}}.
\end{equation}
The corresponding standard deviations are
\begin{equation}
\delta C_{AB}(\boldsymbol{m})=\sqrt{\frac{4\left(m_{++}^{AB}+m_{--}^{AB}\right)\left(m_{+-}^{AB}+m_{-+}^{AB}\right)}{\left(m_{++}^{AB}+m_{--}^{AB}+m_{+-}^{AB}+m_{-+}^{AB}\right)^{3}}}.
\end{equation}
and
\begin{equation}
\delta P_{AU}(\boldsymbol{m})=\sqrt{\frac{(N_{0}-\sum_{BV}m_{UV}^{AB})\sum_{BV}m_{UV}^{AB}}{N_{0}^{3}}},
\end{equation}
as well as

\begin{equation}
\delta D_{BV}(\boldsymbol{m})=\sqrt{\frac{(N_{0}-\sum_{AU}m_{UV}^{AB})\sum_{AU}m_{UV}^{AB}}{N_{0}^{3}}}.
\end{equation}
This gives us a total of 21 constraints derived from the raw detector
counts to be used in the calculation of the secret key fraction.

\selectlanguage{english}%
\bibliographystyle{unsrt}
\bibliography{manuscript7}
\selectlanguage{british}

\end{document}